\documentclass[aps,prb,jmr,twocolumn,notitlepage,superscriptaddress,longbibliography,floatfix,amsmath]{revtex4-1}
\usepackage{color,graphicx}
\usepackage{amssymb}
\usepackage{stmaryrd}
\input{amssym.def}

\begin{document}

\title{Nonequilibrium grain size distribution with generalized growth and nucleation rates}
\author{Kimberly S.~Lokovic$^1$, Ralf B.~Bergmann$^2$, and Andreas Bill}
\email[Author to whom correspondence should be addressed: ]{abill@csulb.edu}
\affiliation{California State University Long Beach, Department of Physics \& Astronomy, 1250 Bellflower Blvd., Long Beach, CA 90840\\
$^2$Institute for Applied Beam Technology (BIAS), Klagenfurter Str.~2, 28359 Bremen, Germany}
\date{J.~Mater.~Res.~{\bf 28}, 1407 (2013)} 

\begin{abstract}
We determine the non-equilibrium grain size distribution during the crystallization of a solid in $d$ dimensions at fixed thermodynamic conditions, for the random nucleation and growth model, and in absence of grain coalescence. Two distinct generalizations of the theory established earlier are considered. A closed analytic expression of the grain size distribution useful for experimental studies is derived for anisotropic growth rates. The main difference from the isotropic growth case is the appearance of a constant prefactor in the distribution. The second generalization considers a Gaussian source term: nuclei are stable when their volume is within a finite range determined by the thermodynamics of the crystallization process. The numerical results show that this generalization does not change the qualitative picture of our previous study. The generalization only affects quantitatively the early stage of crystallization, when nucleation is dominant. The remarkable result of these major generalizations is that the non-equilibrium grain size distribution is robust against anisotropic growth of grains and fluctuations of nuclei sizes.
\end{abstract}


\maketitle

%%%%%%%%%%%%%%%%%%%%%%%%%%%%%%
\section{Introduction}\label{s:intro}

\begin{figure}[h] 
\begin{center}
\includegraphics[scale=0.55]{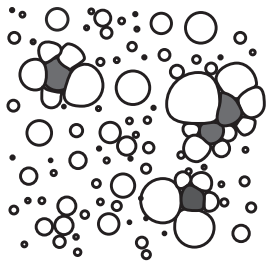}
\end{center}
\caption[Grain distribution at intermediate stages of crystallization]{Schematic representation of the grain distribution at intermediate stages of crystallization. The smallest dots are nuclei created randomly in space and time. The dark grains are completely surrounded by other grains and have thus stopped growing because of impingement, absence of coalescence and secondary growth. In this figure, unimpinged grains are disk like, corresponding to an isotropic growth rate. Impingement changes the shape of grains. In Sec.~\ref{s:anisotropic} we generalize the theory to anisotropic growth rates.}
\label{fig:tessellation}
\end{figure}

Electron microscope images done on annealed amorphous materials generally reveal a tessellation of crystalline grains with a distribution of sizes and shapes. The inhomogeneous end product of the crystallization process has its origin in the formation of nuclei and their growth into grains as schematically shown in Fig.~\ref{fig:tessellation}.

These can occur in a variety of ways, one of which is the random formation of stable clusters with respect to thermodynamic fluctuations within the sample, and the growth of these nuclei into grains. An important example of such random nucleation and growth (RNG) process is observed in the crystallization of silicon thin films used for solar cells (see, {\it e.g.}, Refs.~\onlinecite{wagnerPSSA10,*HandbookPhotovoltaicv11} and references therein). Characterizing quantitatively the grain size distribution (GSD) is important since many physical properties directly depend on this distribution. For example, electrical, magnetic, optical or even superconducting properties are affected by the granularity of a sample.

We have developed recently a theory for the non-equilibrium GSD during the crystallization of a solid in $d$ dimensions\cite{bergmannJCG08,teranPRB10} and applied the theory successfully to the solid phase crystallization of Silicon.\cite{billMRS09} Theoretical studies discussing the grain size distribution in various contexts use analytical tools mainly in one dimension,\cite{gelbardJCIS79,shiJMR91,shiJMR91b,shiMCP94,sekimotoPA84,sekimotoIJMPB91,axePRB86,bennaimPRE96,junPRE05} or numerical techniques\cite{brownJAP95,fayadSM99,riosSM99,wangISIJI03,axePRB86,crespoPRB96,pinedaPRB99,pinedaPRE04,brunaJAP06,soderlundPRL98,farjasPRB07,farjasPRB08} to describe the formation of grains during crystallization. The time-dependent GSD derived in Refs.~\onlinecite{bergmannJCG08,teranPRB10} and extended here is obtained analytically for $d$ dimensions as a solution of a differential equation that accounts for the random creation of nuclei over the volume of the sample and their growth into grains. Besides the ability to describe the distribution during the crystallization process, one other important outcome of the theory is that the lognormal distribution of grain sizes generally postulated to fit experimental data at full crystallization results naturally from our theory when the effective nucleation {\it and} growth rates are time-decaying quantities.\cite{bergmannJCG08,teranPRB10} As discussed in Ref.~\onlinecite{teranPRB10}, the proposed differential equation for the distribution is rather general and the theory may thus apply to a variety of phenomena not related to crystallization.

The main purpose of this paper is to extend the theory in two major ways and demonstrate that the non-equilibrium GSD derived previously is robust under various generalizations. We also provide an expression for the GSD (Eq.~\eqref{solanistropic} below) that can be easily used to analyze experimental data.  The first generalization  allows for {\it anisotropic} grain growth. In Refs.~\onlinecite{bergmannJCG08,teranPRB10} we had assumed that unimpinged grains remain spherical (or disk-like in 2D thin films). Yet, it is known experimentally that unimpinged grains can display many different shapes such a ellipsoidal or filamentary (see {\it e.g.} Ref.~\onlinecite{kakinumaJVSTA95,*hartman73,*gentryPRB09}). Such non-spherical shapes result from the anisotropic growth of grains. In the second extension of the theory we introduce a nucleation rate that accounts for the fact that local thermodynamic and structural fluctuations allow clusters of various sizes to become stable nuclei. Previously we had assumed that only nuclei of a well defined size are formed. We show below that the theory is robust against relaxing this condition to allow for the formation of nuclei with sizes varying within an experimentally reasonable range of a given average.

The theory presented in Refs.~\onlinecite{bergmannJCG08,teranPRB10} applies when nucleation occurs randomly over the sample at a constant microscopic rate $I_0$ and nuclei grow into grains at constant microscopic rate $v_0$. We assume that grains do not coalesce to form larger grains. We also assume the absence of secondary grain growth.  Our assumptions imply that the total number of grains per unit volume of the sample, $N(t)$ is a monotonously increasing function of time; at fixed number of nuclei the crystallization process consists in the growth of a constant number of grains. The distribution of grain sizes then obeys a continuity equation. Since we assume simultaneous nucleation and growth of grains the differential equation must include a source term for random nucleation. The resulting equation for the non-equilibrium GSD $N(\mathbf{r},t)$ is\cite{bergmannJCG08,teranPRB10}
\begin{eqnarray}
\label{PDE}
\frac{\partial N(\mathbf{r},t)}{\partial t} + \boldsymbol{\nabla}_{\mathbf{r}}\cdot\left[N(\mathbf{r},t)\,\mathbf{v}(\mathbf{r},t)\right] = \mathcal{D}(\mathbf{r},t)\,,
\end{eqnarray}
where the right hand side of the equation is the source term for nucleation and $\mathbf{v}$ is the general growth rate of grains. The source term is written as a product $\mathcal{D}(\mathbf{r},t) = J(t) D(\mathbf{r})$, where $J(t)$ is the time dependent effective nucleation rate and $D(\mathbf{r})$ is the source term of nuclei. Both quantities are defined below. $\mathbf{r} = (r_1,\dots,r_d)$ is a $d-$dimensional vector characterizing the size of a grain. As discussed in the next section we describe grains as $d-$dimensional ellipsoids and the components $r_i$ ($i=1,\dots,d$) of the vector are the semi-axes of the ellipsoid. For isotropic growth rates\cite{teranPRB10} we have $r_1=\dots =r_d=\rho$, the radius of the $d-$dimensional sphere. Other possible definitions of $\mathbf{r}$ are suggested in Ref.~\onlinecite{teranPRB10}. In section \ref{s:anisotropic} we assume $r_1 > r_2 \geq r_3$ and in Sec.~\ref{s:gaussian} we take $r_i=\rho$ ($i=1,\dots,d$).

Solving Eq.~\eqref{PDE} requires the knowledge of the effective nucleation and growth rates, $J(t)$ and $\mathbf{v}(\mathbf{r},t)$, respectively. It is important to realize that though the microscopic rates $I_0$ and $v_0$ are constant, the {\it effective} rates are time-dependent. Assuming that no nuclei can form in the volume of an existing grain, the probability of formation of a new nucleus must decay in time since the volume available for its formation decreases in time. Kolmogorov, Mehl, Johnson and Avrami (KMJA) have derived an exact expression for the effective nucleation rate $J(t)$ in the case of RNG processes in $d$ dimensions\cite{kolmogorov37,mehljohnson,avrami39,*avrami40,*avrami41}
\begin{eqnarray}\label{Jt}
J(t) = I_0 e^{-\left(\frac{t-t_0}{t_{cI}}\right)^{d+1}}\Theta\left(\frac{t-t_o}{t_{cI}}\right) \equiv I_0 f(t).
\end{eqnarray}
The expression for $f(t)$ is well-established as discussed in numerous publications (see references in Ref.~\onlinecite{teranPRB10}) and applies to RNG processes. $t_0$ is the incubation time, $t_{cI} = \left[(d+1)/I_0v_0\omega_d\right]^{1/d+1}$ is the critical time for nucleation and $\Theta$ is the Heaviside function. $\omega_d$ is the constant appearing in the volume $\Omega_d = \omega_d r_1\cdots r_d$ of the hypersphere.\cite{teranPRB10}  For example for $d=1,2,3$ we have $\omega_1 = 2$, $\omega_2 = \pi$, $\omega_3 = 4\pi/3$, respectively.

The effective growth rate similarly decreases in time since impingement stops the growth of a grain in the direction perpendicular to the contact line between grains. Once completely surrounded by other grains a given grain cannot grow anymore (see Fig.~\ref{fig:tessellation}). The time-dependence of the effective growth rate is postulated to have a form similar to the one for $J(t)$. The  discussion of Refs.~\onlinecite{bergmannJCG08,teranPRB10,billMRS09} lead to consider an exponential time decay of the effective growth rate
\begin{eqnarray}\label{vt}
v(t) = v_0 e^{-\left(\frac{t-t_0}{t_{cv}}\right)} \Theta\left(\frac{t-t_0}{t_{cv}}\right) \equiv v_0 g(t)\,.
\end{eqnarray}
Both the microscopic growth rate $v_0$ and the critical time $t_{cv}$ can be determined experimentally.\cite{bergmannJCG08}

Finally, to solve Eq.~\eqref{PDE} we need an expression for the source term of nuclei, $D(\mathbf{r})$. We consider two cases. In one case, we assume that only nuclei of a specific critical volume $\Omega_c$ can form. In this case, $D(\mathbf{r}) = \delta\left(\Omega - \Omega_c\right)$ is given by a Dirac distribution. This assumption allows for an analytic treatment of the equation\cite{bergmannJCG08,teranPRB10} and will be considered in the next section, Sec.~\ref{s:anisotropic}, dealing with anisotropic growth rates. In Sec.~\ref{s:gaussian} we generalize the theory by relaxing this condition. We consider the more physical standpoint according to which a cluster is thermodynamically stable when within a range of volumes around a mean value. We consider the case of a Gaussian distribution
\begin{eqnarray}\label{DrgaussianOm}
D(\mathbf{r}) = \frac{1}{\epsilon\,\sqrt{2\,\pi}}\,e^{-\frac{1}{2\epsilon^2}(\Omega - \Omega_c)^2},
\end{eqnarray}
where $\epsilon$ is a small real number and $\Omega_c$ is now the mean value of a nucleus' volume. The study of the nucleation barrier that has to be overcome by a cluster of atoms or molecules and stabilizes the nucleus has been studied in detail in Ref.~\onlinecite{shiJMR91,*shiJMR91b,*shiMCP94}. Their study gives an expression for $\epsilon$ that can be used to estimate its value for specific systems.

%%%%%%%%%%%%%%%%%%%%%%%%%%%%%%
\section{Anisotropic growth rate}\label{s:anisotropic}

\begin{figure*} 
\begin{center}
\includegraphics[scale=0.5]{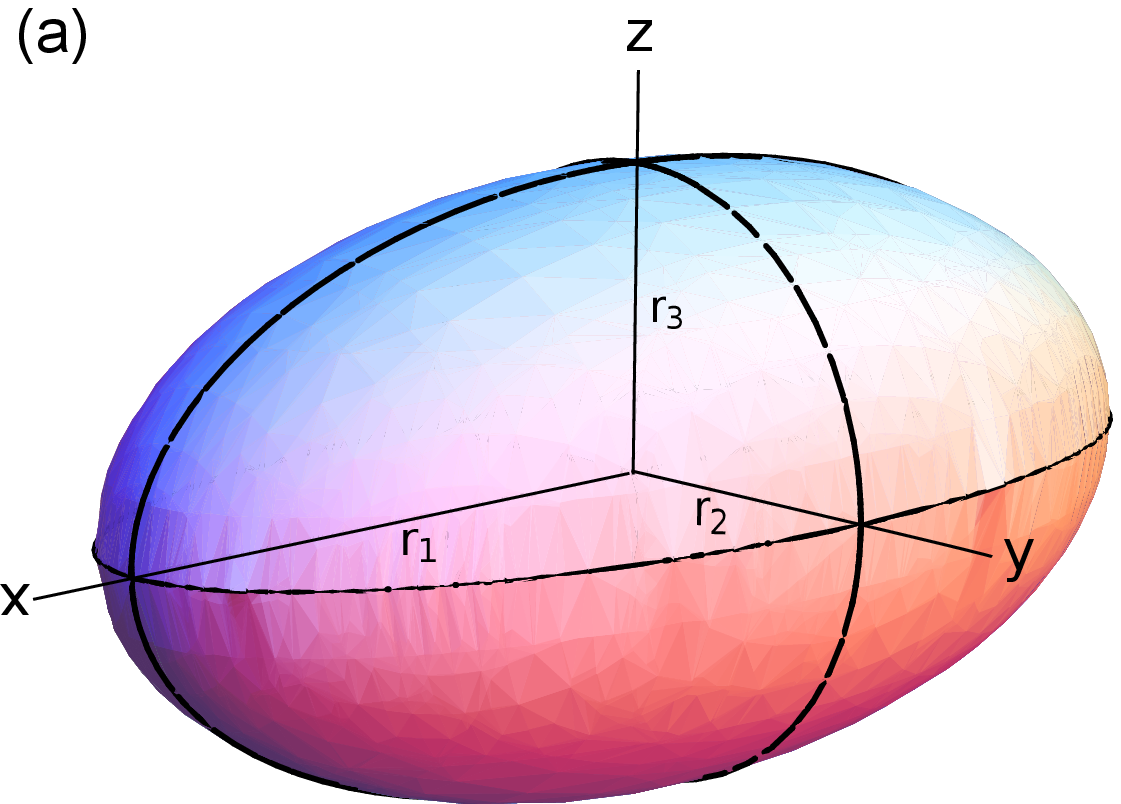}
\includegraphics[scale=0.5]{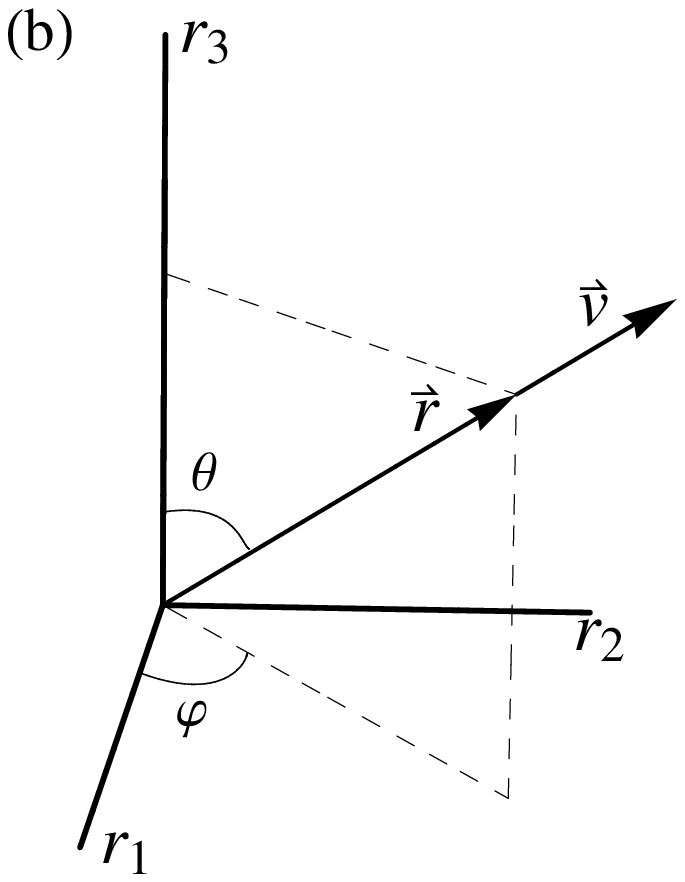}
\end{center}
\caption[Anisotropic grain and space of grain sizes]{(a) Ellipsoidal grains resulting from an anisotropic growth rate (see text). (b) Space of grain sizes spanned by the semi-axes of the ellipsoid. The non-equilibrium GSD $N(\mathbf{r},t)$ is calculated in this space. The growth rate $\mathbf{v}$ points along a radial line in that space since we assume that the anisotropic growth changes the volume of the grain but keeps its shape invariant (see text).}
\label{fig:ellipsoid}
\end{figure*}

In the theory developed in Refs.~\onlinecite{bergmannJCG08,teranPRB10,billMRS09} we assumed that the growth of grains is isotropic, leading to spherical grains when they do not impinge on each other. This assumption is not always valid as even isolated grains often appear non-spherical in shape. The shape is determined by either extrinsic or intrinsic factors. A patterned substrate containing steps may lead to filamentary grains and is an example of an extrinsic factor. On the other hand, when the interaction between atoms or molecules is directional, the grain may also become non-spherical. For example, in Silicon the growth rate is different along different principal axes of the crystal.\cite{kakinumaJVSTA95,*hartman73} Another example is the case of planar molecules that often experience a strong binding when stacked, but only weakly interact with each other through the edges of the molecules.\cite{*gentryPRB09} Although thermodynamic fluctuations will favor the stacking, the binding through the edges may not be completely neglected, leading to an ellipsoidal shape of grains (see Fig.~\ref{fig:ellipsoid}).\\
In compounds made of atoms rather than molecules ({\it e.g.} polycrystalline Si) the formation of  periodic bond chains and the mobility of atoms determine the crystal axis along which the growth preferably occurs.\cite{hartman73} Hence, interactions between the entities constitute an intrinsic factor leading to anisotropic growth of grains. It is important to note that while extrinsic factors generally impose the growth of grains along specific directions of space, the intrinsic factors allow for a random orientation of the ellipsoids in the volume of the sample. Furthermore, since microscopic interactions between molecules are responsible for the shape of the grain, the growth of a grain will mainly consist of an increase in volume and keep the shape invariant. That is, the ratio of semi-axes of the ellipse (for $d=2$) or the ellipsoid ($d=3$) is constant: $r_j/r_1 = r_{cj}/r_{c1}$ for $j=2,3$, $r_1$ ($r_{c1}$) being the major semi-axis of the grain (nucleus). In spherical coordinates it means that $\varphi$ and $\theta$ are constant in $\mathbf{r}$-space (see Fig.~\ref{fig:ellipsoid}). The calculations performed below rely on these physical assumptions.  When considering anisotropic growth we will assume it is due to intrinsic factors. We also assume the absence of nucleation centers in the amorphous sample, and thus a homogeneous and isotropic nucleation rate.

The effective growth rate can be written in the form
\begin{eqnarray}\label{vrt}
\mathbf{v}(\mathbf{r}, t)&= v_1(t)\hat{r}_1 + v_2(t)\hat{r}_2 + v_3(t)\hat{r}_3 = v(t)\,\hat{r},
\end{eqnarray}
where the last equality is obtained by introducing spherical coordinates in $\mathbf{r}$-space and using the fact mentioned above that the shape of the ellipsoid does not change as the grain grows. $v(t)$ is given by Eq.~\eqref{vt}. We emphasize that the vector is written in the space of semi-axes of the ellipsoid $(r_1,r_2,r_3)$ and not in real space $(x,y,z)$. This expression for $\mathbf{v}(\mathbf{r},t)$ is valid in absence of diffusion of atoms during the crystallization process as is for example the case in solid phase crystallization.\cite{bergmannJCG08} We also write the nucleation source term in polar (spherical) coordinates
\begin{eqnarray}\label{Drdelta}
D(\mathbf{r}) = \delta(\Omega_d - \Omega_{c,d}) = \frac{1}{A_{c,d}}\delta(r-r_c)
\end{eqnarray}
where $A_{c,d}$ is the area of the nucleus in $d$ dimensions
\begin{subequations}\label{Acd}
\begin{eqnarray}
A_{c,2} &=& 2\,\pi\,r_c\,\cos\varphi\,\sin\varphi,\\
A_{c,3} &=& 4\,\pi\,r_c^2\,\sin^2\theta \cos\theta \sin\varphi \cos\varphi.
\end{eqnarray}
\end{subequations}

With Eqs.~(\ref{Jt},\ref{vrt}-\ref{Acd}) the partial differential equation, Eq.~\eqref{PDE}, can be solved analytically for the anisotropic case. The calculation is more involved than for the isotropic growth rate but follows the procedure presented in Ref.~\onlinecite{teranPRB10}. Hence, we do not repeat the derivation here. The general result for the non-equilibrium GSD with anisotropic growth rate is
\begin{widetext}
\begin{eqnarray}\label{solanistropic}
N(\gamma,\tau) &=& \left(\frac{I_o}{v_0}\right) \frac{1}{A_{\infty,d}}\,\gamma^{d-1}\,\frac{f[\sigma(\gamma,\tau)]}{g[\sigma(\gamma,\tau)]}%\,\nonumber\\ &&\times
\left\{\Theta\left(\frac{\gamma-\gamma_c}{1-\gamma_c}\right) - \Theta\left[\frac{\gamma-\gamma_{max}(\tau)}{1-\gamma_c}\right]\right\},
\end{eqnarray}
\end{widetext}
where $f$ and $g$ are given by Eqs.~(\ref{Jt},\ref{vt}) with time replaced by $\sigma(\gamma,\tau)$ below.  This expression is written in terms of the dimensionless quantities
\begin{eqnarray} \label{dim}
\gamma=\frac{r}{r_\infty},\quad \tau=\frac{t}{\sqrt{t_{cv}t_{cI}}},
\end{eqnarray}
where $r_\infty$ is the magnitude of $\mathbf{r}$ for the largest grain found at full crystallization, $\gamma_c = r_c/r_\infty$, and $\sigma(t)$ is given by\cite{teranPRB10}
\begin{eqnarray}\label{sigmart}
\sigma(\gamma,\tau) = \tau_0 + t_r\,\ln\left( \frac{\gamma-\gamma_c}{\mathcal{V}_0} + e^{-(\tau-\tau_0)/t_r} \right)^{-1},
\end{eqnarray}
with $t_r^2 = t_{cv}/t_{cI}$ and ${\mathcal V}_0 = t_{cv}v_0/r_\infty$.

The central result, Eq.~\eqref{solanistropic}, is a generalization of Eq.~(21) in Ref.~\onlinecite{teranPRB10} to the case of anisotropic growth rates.  It can be implemented in a variety of ways to describe experimental data or extract specific parameters of the model by fitting to experimental findings.\cite{billMRS09}  The result has several interesting features we now describe.

The term in curly parenthesis is easily understood. It states that the non-equilibrium GSD $N(\gamma,\tau)$ is only non-zero in the interval $\gamma_c\leq \gamma \leq \gamma_{\rm max}$, that is, for grain sizes between those of a nucleus ($\gamma_c$) and of the largest grain found at time $t$ in the sample [$\lim_{\tau\to\infty} \gamma_{\rm max}(\tau) = 1$ since $r_{\rm max}(t\to\infty) = r_\infty$)].
Further, the non-equilibrium GSD is proportional to the ratio of microscopic nucleation and growth rates $I_0/v_0$. It is also proportional to the rate $f(\sigma)/g(\sigma)$. Note, however, that since $\sigma(\gamma,\tau)$ is a non-linear function of time and grain size, the ratio is {\it not} simply the ratio of Eqs.~\eqref{Jt} and \eqref{vt}. Finally, the non-equilibrium GSD is inversely proportional to the surface area $A_{\infty,d}$ of the largest grain found at full crystallization. The expression for $A_{\infty,d}$ is the same as in Eq.~\eqref{Acd} with $r_c$ replaced by $r_\infty$.

Eq.~\eqref{solanistropic} is a remarkable result. It states that the only difference between isotropic and anisotropic growth rates is the presence of $A_{\infty,d}$ in the prefactor. Thus, the results derived in Refs.~\onlinecite{bergmannJCG08,teranPRB10} are robust against the generalization to anisotropic growth rates!

\begin{figure}[htb] 
\begin{center}
\includegraphics[scale=0.55]{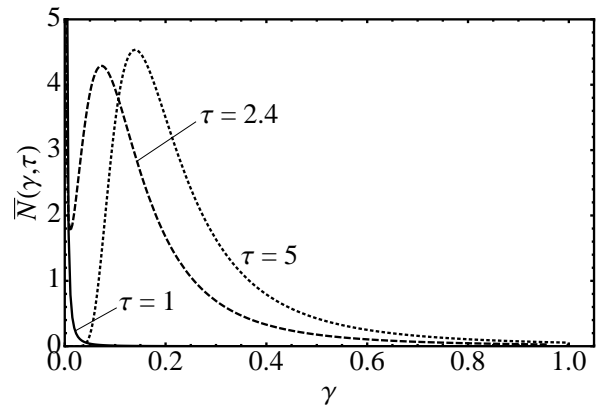}
\end{center}
\caption[Normalized GSD, anisotropic growth, $\tau$ and $d=3$]{Normalized grain size distribution in the early stage ($	\tau = 1$), intermediate stage ($\tau=2.4$) and late stage ($\tau=5$) of crystallization for $d=3$. In the early stage only the 	nucleation peaks is observed. At intermediate stage the nucleation peak appears concomitantly with the broader growth 	peak. At late stages of crystallization, nucleation has disappeared and only the growth peak remains. The curve at full 	crystallization $\tau \geq 5$ is lognormal like.\cite{teranPRB10} Parameters are $\gamma_c = 0.0001, \tau_0 = 0, t_r = 	0.75, \mathcal{V}_0 = 1$ (see Refs.~\onlinecite{bergmannJCG08,billMRS09}).}
\label{fig:GSDtau}
\end{figure}

Fig.~\ref{fig:GSDtau} displays the normalized non-equilibrium GSD $\bar{N}(\gamma,\tau) = N(\gamma,\tau)/N(\tau)$, where
\begin{eqnarray}\label{Nt}
N(\tau) = \int_{\gamma_c}^{\gamma_{\rm max}(\tau)} N(\gamma,\tau')\,d\tau',
\end{eqnarray}
for early, intermediate and late stages of crystallization\cite{teranPRB10} in three dimensions ($d=3$).
In the early stage, only a nucleation peak appears as the initial process is dominated by the formation of nuclei. The intermediate stage is characterized by the simultaneous appearance of a nucleation peak (near $\gamma_c$) and a growth peak. That stage occurs over a very short period; in the present calculation, full crystallization is reached for $\tau \gtrsim 5$ and the time interval of the intermediate stage is $2\lesssim \tau \lesssim 3$. The late stage of crystallization is obtained for $\tau \gtrsim 3$ where nucleation is almost completely suppressed and grains continue growing in areas of untransformed material until complete tessellation of the sample is achieved. Thus, in the late stage of crystallization the shape of the distribution remains unchanged but shifts towards larger grain sizes. As discussed in Refs.~\onlinecite{bergmannJCG08,teranPRB10} the GSD for $\tau \geq 5$ in Fig.~\ref{fig:GSDtau} displays a lognormal like distribution.
The results of the GSD for $d=2$, that is for a thin film where the average grain size at full crystallization is larger than the thickness of the film, are qualitatively similar to the $d=3$ case depicted in Fig.~\ref{fig:GSDtau} and is therefore not shown here (see Ref.~\onlinecite{teranPRB10}). 

The remarkable conclusion of this section is that the non-equilibrium GSD for isotropic and anisotropic growth rates are qualitatively similar as they only differ by the constant prefactor $A_{\infty,d}^{-1}$.

%%%%%%%%%%%%%%%%%%%%%%%%%%%%%%
\section{Gaussian nucleation rate}\label{s:gaussian}

The theory developed in Refs.~\onlinecite{bergmannJCG08,teranPRB10,billMRS09}  and used in the previous section considered the  simplified expression, Eq.~\eqref{Drdelta}, for the source term of nuclei $D(\mathbf{r})$. We assumed that only nuclei of a well-defined specific volume $\Omega_c$ are stable in the system. Such assumption is physically not  entirely realistic as grains with a few more or a few less molecules are likely to be thermodynamically stable as well. In the present section we analyze how the non-equilibrium GSD is affected if one relaxes the condition imposed by Eq.~\eqref{Drdelta}.  Contrary to the previous section but in accordance with Refs.~\onlinecite{bergmannJCG08,teranPRB10,billMRS09} we assume isotropic growth of grains, leading to spherically shaped unimpinged grains.

To generalize the theory we assume that the Dirac distribution appearing in Eq.~\eqref{Drdelta} is replaced by the Gaussian distribution, Eq.~\eqref{DrgaussianOm}. In dimensionless quantities and using spherical coordinates $D(\mathbf{r})$ reads
\begin{eqnarray}
D(\gamma) = \frac{1}{r_\infty\, \epsilon'\,\sqrt{2\,\pi}}\,e^{-\frac{1}{2\,\epsilon'^2}(\gamma - \gamma_c)^2}\,
\end{eqnarray}
where $\epsilon' \equiv \epsilon/r_\infty$. This term has to be inserted into the right hand side of Eq.~\eqref{PDE}. Contrary to all previous calculations the present form of the differential equation allows solving for the non-equilibrium GSD analytically except for one integral. Defining
\begin{eqnarray}\label{Ntilde}
\tilde{N}(\gamma,\tau) = \gamma^{d-1}\,A_{\infty, d}\,r_\infty^2\,N(\gamma,\tau),
\end{eqnarray}
the normalized GSD for a Gaussian nucleation rate becomes
\begin{widetext}
\begin{eqnarray}\label{solgaussian}
\tilde{N}(\gamma,\tau) &=& \frac{2\,\epsilon'\,t_r^2}{\gamma_c\sqrt{2\,\pi}}e^{-\frac{\gamma_c^2}{2\,\epsilon'^2}}\,\frac{I(\sigma)}{v(\sigma)}\,\delta_{d,2}%\nonumber\\
%&&
\mp \frac{t_r}{2\,\gamma_c^{d-1}}\int_{\tau_0}^\tau \frac{I(\tau')}{\epsilon' \sqrt{2\,\pi}}\left[\gamma - u(\tau',\tau)\right]^{d-1}%\nonumber\\ && \hspace*{0.4cm} \times \,\,
\exp\left\{-\frac{\left[\gamma \mp \gamma_c - u(\tau',\tau)\right]^2}{2\,\epsilon'^2}\right\}d\tau',
\end{eqnarray}
\end{widetext}
where the minus (plus) sign is for $d=2$ ($d=3$), Eq.~\eqref{sigmart} defines $\sigma$, and
\begin{eqnarray}
u(\tau',\tau) = \mathcal{V}_0 \left[\exp\left(-\frac{\tau-\tau_0}{t_r}\right) - \exp\left(-\frac{\tau'-\tau_0}{t_r}\right)\right].
\end{eqnarray}
Because of the Kronecker symbol $\delta_{d,2}$ the first term is absent for $d=3$. That term vanishes in the limit $\epsilon\to 0$, in which case the Gaussian in the second term reduces to the Dirac distribution and the integral can be solved, leading to the result previously established in Refs.~\onlinecite{bergmannJCG08,teranPRB10}.

Since for $\epsilon\neq 0$ the above integral cannot be solved analytically we performed numerical calculations to study the time dependent form of the GSD. To comply with typical experimental results (see, {\it e.g.}, Ref.~\onlinecite{shiJMR91,*shiJMR91b,*shiMCP94}) we assumed that the width of the Gaussian is of the order of a tenth of the average nuclei radius, $\epsilon\sim 0.1 r_c$. The results for the normalized distribution differ from those of the previous section represented in Fig.~\ref{fig:GSDtau} only for early stages of crystallization, and only marginally. The nucleation peak is slightly wider in the early and intermediate stages. Hence, Eq.~\eqref{solgaussian} should only be used to discuss experimental studies of early stages of nucleation. For all other cases, a nucleation source term with a Dirac distribution is sufficient.

%%%%%%%%%%%%%%%%%%%%%%%%%%%%%%
\section{Conclusion}\label{s:conclusion}
We studied two major generalizations of the theory established in Refs.~\onlinecite{bergmannJCG08,teranPRB10} to determine the non-equilibrium grain size distribution during the crystallization of a solid. In the first, we assumed that the growth of grains can be anisotropic, as observed in some experiments. We showed that the final non-equilibrium grain size distribution remains essentially unaffected by this generalization [except for a constant prefactor in $N(\mathbf{r},t)$]. In the second extension of the theory we considered the case where nuclei with variable size within a physically realistic range can be thermodynamically stable and develop into grains. Using a Gaussian distribution to model the source of nuclei, the partial differential equation could not be completely solved analytically. The numerical results showed that only for early stages of crystallization, when nucleation dominates the process, did the generalization affect the grain size distribution quantitatively. The main conclusion of this work is that the theory established in Refs.~\onlinecite{bergmannJCG08,teranPRB10} is very robust against these important generalizations. This is particularly surprising for the generalization to anisotropic grain growth. This is explained in part by the fact that we assume the shape of unimpinged grains to remain unaltered during the growth.  This  assumption of our model is physically justified by the fact that the formation of a nucleus and its growth into a grain are determined by the microscopic interactions between atoms and molecules, the intrinsic factors described in the introduction.

In both generalizations we found the previously obtained behavior of the non-equilibrium grain size distribution, which goes through three stages during the crystallization: a nucleation dominated early stage, an intermediate stage where nucleation and growth have similar strength and a late stage of crystallization where growth dominates.\cite{teranPRB10} We also note that in all cases studied the non-equilibrium grain size distribution is found to depend on the ratio of the effective nucleation and growth rates, $N(r,t) \propto I[\sigma(r,t)]/v[\sigma(r,t)]$ with, however, a non-trivial function $\sigma(r,t)$ of grain size and time.  The assumptions leading to that result are the random nucleation and growth of grains at fixed thermodynamic conditions and in the absence of coalescence or secondary grain formation.  When using our expression to describe the crystallization of specific materials it is important to carefully consider the latter assumption since coalescence may affect the grain size distribution.
 In conclusion, Eq.~\eqref{solanistropic} provides an excellent description of the non-equilibrium grain size distribution in systems where random nucleation and growth occurs and can be used to analyze experimental data.

\begin{acknowledgements}
We gratefully acknowledge the support of the Research Corporation, the Army Research Laboratory and the CNSM block grant at CSU Long Beach.
\end{acknowledgements}

\newpage
\bibliography{Lokovic_JMR_BiBTeX}

\end{document}